\documentclass[12pt,preprint]{aastex}

\begin{document}

\title{V605 Aql: The Older Twin of Sakurai's Object\altaffilmark{1}}

\author{ Geoffrey C. Clayton\altaffilmark{2,3}, F. Kerber\altaffilmark{4}, N. Pirzkal\altaffilmark{5}, O. De Marco\altaffilmark{6}, P.A. Crowther\altaffilmark{7}, and J.M. Fedrow\altaffilmark{3,8}}

\altaffiltext{1}{Based on observations collected at the European Southern Observatory, Chile, Proposal 67.D-0405}
\altaffiltext{2}{Department of Physics \& Astronomy, Louisiana State
University, Baton Rouge, LA 70803; gclayton@fenway.phys.lsu.edu}
\altaffiltext{3}{Maria Mitchell Observatory, Maria Mitchell Association, 3 
Vestal Street, Nantucket MA 02554}
\altaffiltext{4}{European Southern Observatory, Karl-Schwarzschild-Stra§e 2, D-85748 Garching, Germany; fkerber@eso.org}
\altaffiltext{5}{STScI, 3700 San Martin Dr., Baltimore, MD 21218; npirzkal@stsci.edu}
\altaffiltext{6}{Department of Astrophysics, American Museum of Natural History, Central Park West at 79th Street, New York, NY 10024; orsola@amnh.org}
\altaffiltext{7}{Department of Physics \& Astronomy, University of Sheffield, Hicks Building, Hounsfield Rd, Sheffield, S3 7RH, UK; paul.crowther@sheffield.ac.uk}
\altaffiltext{8}{The Evergreen State College, 2700 Evergreen Parkway NW, Olympia, Washington 98505}

\begin{abstract}

New optical spectra have been obtained with VLT/FORS2 of the final helium shell flash (FF) star, V605 Aql,  which peaked in brightness in 1919.
New models suggest that this star is experiencing a very late thermal pulse.  The evolution to a cool luminous giant and then back to a compact hot star takes place in only a few years. 
V605 Aql, the central star of the Planetary Nebula (PN), A58,  has evolved from T$_{eff}\sim$5000 K in 1921 to $\sim$95,000 K today. There are indications that the new FF star, Sakurai's Object (V4334 Sgr), which appeared in 1996, is evolving along a similar path. The abundances of Sakurai's Object today and V605 Aql 80 years ago mimic the hydrogen deficient R Coronae Borealis (RCB) stars with 98\% He and 1\% C. The new spectra show that V605 Aql has stellar abundances similar to those seen in Wolf-Rayet [WC] central stars of PNe with $\sim$55\% He, and $\sim$40\% C. 
The stellar spectrum of V605 Aql can be seen even though the star is not directly detected. Therefore, we may be seeing the spectrum in light scattered around the edge of a thick torus of dust seen edge-on. 
In the present state of evolution of V605 Aql, we may be seeing the not too distant future of Sakurai's Object. 
\end{abstract}


\section{Introduction}
 
In 1996, Sakurai's Object (V4334 Sgr) was discovered undergoing a nova--like outburst  
 \citep{1996IAUC.6322....1N, 1996IAUC.6325....1B, 1996IAUC.6328....1D}. Subsequent observations showed Sakurai's Object to be a star at the center of an old Planetary 
Nebula (PN) undergoing a final helium shell flash (FF) \citep{1983ApJ...264..605I, 2001Ap&SS.275....1B}.
In 1919, a very similar object, V605 Aquilae, was discovered which was initially thought to be a slow nova \citep{1920AN....211..119W}.
Only one other FF object has been discovered while in outburst. This is FG Sge, which is evolving much more slowly \citep{1968ApJ...153..397H, 1998ApJS..114..133G}.
While FG Sge seems to be evolving very differently, V605 Aql and Sakurai's Object seem to be twins~\citep{2002Ap&SS.279....5D,2004A&A...426..145L}.
Understanding the FF stars is a key test
for any theory which aims to explain the evolution of
post-AGB stars and hydrogen deficiency.

V605 Aql brightened over a period of two years to a peak of 
$m_{pg}$=10.2 in 1919  \citep{1921PASP...33..314L}.
Its spectrum in 1921 was very similar to an R Coronae 
Borealis (RCB) star with $T_{eff}\sim$~5000 K \citep{1997AJ....114.2679C,1996PASP..108..225C}. 
Over the period 1919--1923, V605 Aql underwent three 
episodes 
of fading and brightening  \citep{1985MitAG..63..181S, 1996PASP..108.1112H}.
It was 
observed until 1923 at which time it disappeared from the sky and from the published 
literature.
The position of V605 Aql was found to coincide with the 
center of the 
PN, A58 \citep{1971ApJ...170..547F}. The PN has a small hydrogen deficient knot of nebulosity located at its geometric 
center \citep{1985MitAG..63..181S,1996ApJ...472..711G}.

Spectroscopy of the central knot,
obtained in 1986--87, shows a broad emission feature at $\sim$5800 \AA, identified as C~{\sc iv}, and believed to be of stellar origin \citep{1987MitAG..68..244S,1987Msngr..50...14S,1996ApJ...472..711G}.
The absence of additional stellar features can be attributed to the extreme faintness of the
star \citep[$m_V\gtrsim 23$]{1997AJ....114.2679C,2002Ap&SS.279...31B}.  The broad C~{\sc iv} emission feature implies
a Wolf--Rayet--type
spectrum for the central star and a T$_{eff}$ in excess of 50,000 K \citep{1997AJ....114.2679C}.  Radio observations indicate that the ionization of V605 Aql has increased significantly since 1987, so perhaps the T$_{eff}$ has continued to increase \citep{2006astro.ph..5156V}.

In this letter, we report new, more sensitive optical spectra of V605 Aql which allow us to estimate the present T$_{eff}$ and abundances of the star. 
 
\section{Observations and Analysis}

New optical spectra of V605 Aql were obtained on 2001 June 16, 18, 20 and 26 using FORS2 on the VLT-UT4 with GRIS\_300V, centered at 5900 \AA, and the GG375 filter. The spectra have exposure times of 2698, 1799, 1147, and 
2698 s, respectively.
Sensitivity curves were determined for each night using 
lamp spectra and observations of standard ESO spectroscopic 
calibrators.
Wavelength polynomial solutions for each of the V605 Aql
observations were derived by extracting and matching calibration 
lamp spectra. Fifth order polynomials were used to minimize the 
errors of the fit to the dispersion relation.
The wavelength calibration is accurate to about 0.5 \AA~after using night sky 
lines for correcting the zero point.
The flux calibration is good to about 10\% for the stronger lines.
An 8-pixel extraction was made, centered on the star. 
The background was fit and subtracted and the four spectra were summed. The final spectrum is shown in Figure 1. 

This spectrum is a great improvement on the spectra of Seitter, and Guerrero \& Manchado obtained in 1986 and 1993, respectively,  which show only one stellar line, C IV $\lambda$ 5801-5812. 
\citet{1987Msngr..50...14S} also identified stellar features of O V, O VI and He II
although these lines are not evident in her published spectra.
In our new spectrum, a number of  broad stellar lines are visible in addition to the narrow emission lines from the PN. As noted previously, there is  very strong emission at C IV $\lambda$5801-5812. Also strongly present are C IV $\lambda$4658, He II $\lambda$4686, and C IV $\lambda$7724. He II $\lambda$5411, C IV $\lambda$5471, and O V $\lambda$5590 also seem to be weakly present. 

Our spectroscopic analysis utilizes the CMFGEN code \citep{1998ApJ...496..407H} which solves the transfer equation in the comoving 
frame subject to statistical and radiative equilibrium, assuming an 
expanding, spherically symmetric atmosphere, allowing for metal line 
blanketing and clumping. The stellar radius is defined as the inner 
boundary of the model atmosphere and corresponds to Rosseland optical
depth $\sim$20, whilst T$_{eff}$ is defined by the 
usual Stefan-Boltzmann relation. As discussed below, the apparent flux and reddening inferred from the observed spectrum are not indicative of their actual values so they cannot be used as constraints on the absolute stellar luminosity. Therefore, we adopt a value of L = 10$^4$ L$_{\sun}$ predicted by evolutionary models for the FF of V605 Aql. 
This
assumption has negligible influence upon the derived temperature
or elemental abundances.

Our approach follows previous studies \citep[e.g.,][]{2002A&A...392..653C,2006ApJ...636.1033C}
such that diagnostic optical lines of C IV $\lambda$5801-12, and He II $\lambda$4686 plus the
local continuum allow a determination of the stellar temperature,
elemental abundances, mass-loss rate and wind velocity, for an adopted
L = 10$^4$ L$_{\sun}$. The ratio of C IV $\lambda$5801-12/He II 
$\lambda$4686 together with the absence of O VI $\lambda$3811-34 indicate T$_{eff} = 95,000 \pm 10,000$ K ($R_\ast = 0.37\pm 0.07 R_{\sun}$).  A standard 
velocity law, v(r)=v$_{\infty}$(1-R/r$)^{\beta}$, with $\beta$=1.5, reveal a wind 
velocity of 2500 km s$^{-1}$ and mass-loss rate of 1.3 x 10$^{-7}$ M$_{\sun}$~yr$^{-1}$, assuming 
a wind clumping factor of $f$=0.1 (a homogeneous model would be $f^{-1/2}$
higher). In the absence of any identifiable stellar hydrogen features, we assume hydrogen to be absent, and estimate the 
C and O abundances from C/He$\sim$0.25 by number using the optical recombination 
lines of He II, C IV, plus C/O $\sim$10 by number from O V $\lambda$5590, i.e., the mass ratio (\%) is
He:C:O=54:40:5. The uncertainty in the He:C mass is $\sim$10-15\%.

\section{Discussion}
\subsection{A Thick Torus of Dust?}

It seems strange that V605 Aql can be detected spectroscopically but cannot be imaged.
The central feature seen in all images obtained since 1923 is not the star but the expanding central knot of 
hydrogen-deficient material that was ejected as part of the FF event. 
No sign of a stellar PSF is seen in the 1991 Hubble Space Telescope/Faint Object Camera (HST/FOC) image \citep{2002Ap&SS.279...31B}.
In 1991, the knot had a diameter of $\sim$0\farcs5 \citep{2002Ap&SS.279...31B,1997AJ....114.2679C}.

However, \citet{1987MitAG..68..244S} suggested that there is a weak red stellar continuum in the spectrum of the central knot.  A similar red continuum is seen in our new spectra. The observed stellar continuum seems to have significantly less reddening than the star, itself. 
Our models suggest a reddening of only E(B-V)$\sim$1.5 mag is needed in order to fit the observed stellar continuum, assuming no nebular continuum contribution. Whereas, the extinction toward the star is estimated to be E(B-V)$>$3, based on the star's invisibility on images \citep{1997AJ....114.2679C}. 


The observed faintness of V605 Aql could be accounted for if it is still the same absolute luminosity that it was in the 1920's, at a distance of about 3.5 kpc, 
with $A_V >$10 \citep{1997AJ....114.2679C}.  
If $A_V$ is this large, then little or no direct stellar light can reach us. 
But the dust may not have formed in a symmetric shell around the star. 
In V605 Aql, we could be viewing an optically thick torus edge on. 
In this geometry, light can escape perpendicular to the torus and scatter around the edge of the obscuring dust cloud. In this scenario, you would be able to ``see" the stellar spectrum even though the star itself could not be imaged. 
The properties of the expanding knot surrounding V605 Aql are consistent with a bipolar structure with the rear side obscured by dust \citep{1992MNRAS.257P..33P}. Spectroscopic and imaging studies of two older objects (A30, A78) thought to be FF stars, indicate that they may have an equatorial ring and polar nebular knots \citep{1994ApJ...435..722B,1995ApJ...449L.143B,1995ApJ...439..264H}.

\citet{2003RMxAC..15...75K} comments that the lack of scattered blue continuum argues against a torus-like structure. But the situation is quite complicated. 
If we are observing a stellar spectrum in scattered light then neither the measured flux nor the reddening inferred from the spectra will reflect the actual values for V605 Aql. In this case, we are not seeing the light from the star directly transmitted through and reddened by a certain optical depth of dust in the shell. Rather, we are observing light from the star which travels along the poles of the torus where the extinction is significantly less but not zero. This light is then scattered by an unknown distribution of dust above the poles. 
The light will scatter preferentially in the blue, so it will affect the spectral energy distribution in the opposite sense to the reddening. 
The photons scattered in our direction will then pass through and be reddened by interstellar dust (E(B-V)$\sim$0.5 at 3.5 kpc) along the line of sight  \citep{1997AJ....114.2679C}.
Spectropolarimetry of V605 Aql could be used to help determine whether the torus model is correct and to distinguish what fraction of the spectrum is seen in direct and scattered light \citep[e.g.,][]{1992AJ....103.1652W,1993ApJ...417..687W}.

\subsection{The Evolution of V605 Aql}

If we assume that the geometry outlined above is correct, then we have no direct measurement of the absolute luminosity of V605 Aql today. 
However, we can be guided by the early post-FF observations of V605 Aql and Sakurai's object, and by models of stars evolving through a Very Late Thermal Pulse (VLTP).
Both Sakurai's object and V605 Aql are thought to be the result of a VLTP where the star experiences a helium shell flash while on the white dwarf cooling track \citep{2001ApJ...554L..71H,2003ApJ...583..913L,2005Sci...308..231H}.
Observations of V605 Aql and Sakurai's Object are important in testing the predictions of these models concerning convection theory.
The VLTP is required since the evolution of V605 Aql and Sakurai's Object has been so rapid.
In a Late Thermal Pulse which happens on the 
horizontal part of the post-asymptotic giant branch (AGB) track, the return to the AGB takes 100-200 yr which is much too long for Sakurai's Object and V605 Aql \citep{2001ApJ...554L..71H}.
 
Once the FF begins, both helium and hydrogen are  being burned intensely \citep{2001ApJ...554L..71H,2003ApJ...583..913L,2005Sci...308..231H,2006A&A...449..313M}.
The outer hydrogen envelope is being mixed down convectively and burned. But the hydrogen is closer to the surface so it is this energy that causes the first expansion and cooling of the star. It takes the star 5-10 yr to expand and cool back to the AGB. The star then moves back leftward in the HR diagram toward hotter temperatures for 10-50 yr while contracting in size. 
Sakurai's Object has shown an increase in ionization indicating that it may be heating up again after cooling and expanding back to the AGB
\citep{2005Sci...308..231H,2002ApJ...581L..39K}. 
Once there, the star begins a second loop where the energy from the helium burning causes the star to return to the AGB one last time, 250-500 yr later.
During the first loop, the surface abundances of the star undergo large changes. 
A series of high resolution spectra obtained in 1996 show the abundances of Sakurai's Object changing on a timescale of weeks. In particular, the abundance of hydrogen was dropping accompanied by a rise in carbon and s-process elements. \citep{1999A&A...343..507A}. 
It is likely that V605 Aql is finishing the first rapid loop of the VLTP and has evolved more or less horizontally across the HR diagram and today has  L $\sim$ 10$^4$ L$_{\sun}$ similar to its luminosity in the 1920's \citep{2003ApJ...583..913L,2005Sci...308..231H}.
In our spectroscopic analysis in Section 2, we have therefore adopted this luminosity. 
The T$_{eff}$ derived from our spectral synthesis modeling is consistent with this scenario.
These models are constrained by estimates of the properties of the pre-flash progenitor stars using their surrounding PNe  \citep{1999A&A...344L..79K,2004A&A...426..145L,1999MNRAS.304..127P}. 


\subsection{Do R Coronae Borealis Stars evolve from Final Flash stars?}

RCB stars are a rare class of H-deficient supergiants that undergo deep and random brightness declines
due to dust, which forms near the stellar surface and then dissipates (Clayton 1996).
The lightcurve behavior and spectral appearance of V605 Aql in 1921 and the current light curve behavior and
abundances of Sakurai's Object are reminiscient of the RCB class. There are, however, several reasons why
FF stars are unlikely to be the evolutionary precursors of the majority of RCB stars.


The FF objects have deeper light declines ($>$10~mag) than do RCB stars ($\sim$8~mag). The 
abundances of FF objects, shortly after the outburst, do match those of RCB stars, except for one important difference:
the presence of significant amounts of $^{13}C$ in Sakurai's Object, but not in RCB stars. 
In general, an RCB star will have $^{12}C/^{13}C \geq 100$ but in 1996, Sakurai's Object had 1.5 $\lesssim$$^{12}C/^{13}C\lesssim5$ \citep{1994MNRAS.268..544P,1999A&A...343..507A}. The high $^{13}C$ might be a transient feature but measurements in near-IR spectra of Sakurai's object as late as 1998 June find a ratio of $^{12}C/^{13}C = 4\pm1$ 
\citep{1998MNRAS.298L..37E,2004A&A...417L..39P}.
It is possible that the very low resolution 1921 spectrum of V605 Aql cannot resolve the 
$^{12}C^{12}C~\lambda$4737 bandhead from the $^{12}C^{13}C~\lambda$4744 bandhead and so the presence of $^{13}C$ was not noticed \citep{1997AJ....114.2679C}.

For a very short time, perhaps as short as two years, both V605 Aql and Sakurai's Object were almost indistinguishable from the RCB stars  in abundances, temperature, absolute luminosity and lightcurve behavior.
The RCB-like spectrum (and presumably abundance pattern) of V605 Aql in 1921 was already very different in 1987 \citep{1997AJ....114.2679C}. 
Unfortunately, this extremely short RCB phase of the FF stars means that they cannot account for even the small number of RCB stars known in the Galaxy. There are about 50 RCB stars known and it is predicted that there may be as many as $\sim$3000 in the Galaxy as a whole \citep{2001ApJ...554..298A, 2005AJ....130.2293Z}. From R CrB itself, we have a lower limit on the lifetime of an RCB star of $>$200 yr \citep{1996PASP..108..225C}. 

The abundances of V605 Aql in 2001 and Sakurai's Object in 1996 are listed in Table 1, along with the abundances of typical RCB and [WC] stars. 
If all FF stars behave like V605 Aql, their surface compositions will eventually make the transition from mostly helium
to similar fractions of helium and carbon. This transition is likely to be due
to mass-loss which progressively peals  off the helium layer, uncovering the intershell region \citep{2006PASP..118..183W}.
These abundances (helium and carbon in similar amounts) are typical
of the [WC] central stars of PNe (De Marco \& Barlow 2002). These are also the abundances of the two old FF objects
A30 and A78 \citep{2006PASP..118..183W}. These
abundance similarities imply that the FF could have produced the $\sim$60
[WC] central stars known today (Gorny \& Tylenda 2000). But then it is hard to explain why only A30 and A78 show the morphology of a FF object (large H-rich PN surrounding H-poor ejecta).
The properties of V605 Aql today may show the very near future of Sakurai's Object. 
These stars provide a rare opportunity to view a stage of stellar evolution 
which 
proceeds on a timescale of months and years rather than centuries and millennia.

\acknowledgments

This project was supported by the Maria Mitchell Association and NSF/REU grant AST-0354056. PAC acknowledges financial support from the Royal Society.

\clearpage

\begin{deluxetable}{lllll}
\tablewidth{3in}
\tablecaption{Abundances (\%) By Mass}
\tablenum{1}
\tablehead{\colhead{Star} & \colhead{H}& \colhead{He}& \colhead{C}& \colhead{O}}
\startdata
RCB$^a$ star	&0	&98	&1	&0.2\\
$[WC]^a$ star&0&50&40&10\\
Sakurai's Object$^b$&0&90&7&3\\
V605 Aql	&0	&54	&40	&5\\
\enddata
\tablerefs{\citep{2000A&A...353..287A,1996A&A...312..167L,1999A&A...343..507A,2003ApJ...583..913L,2006ApJ...636.1033C,2003IAUS..209..215D, 2003IAUS..209..243C}}
\tablenotetext{a}{Typical abundances for typical RCB stars and  [WC] central stars of PNe.}
\tablenotetext{b}{Abundances on 1996 October 7.}
\end{deluxetable}

\clearpage

\begin{figure}
\figurenum{1}
\includegraphics[scale=0.8,angle=-90]{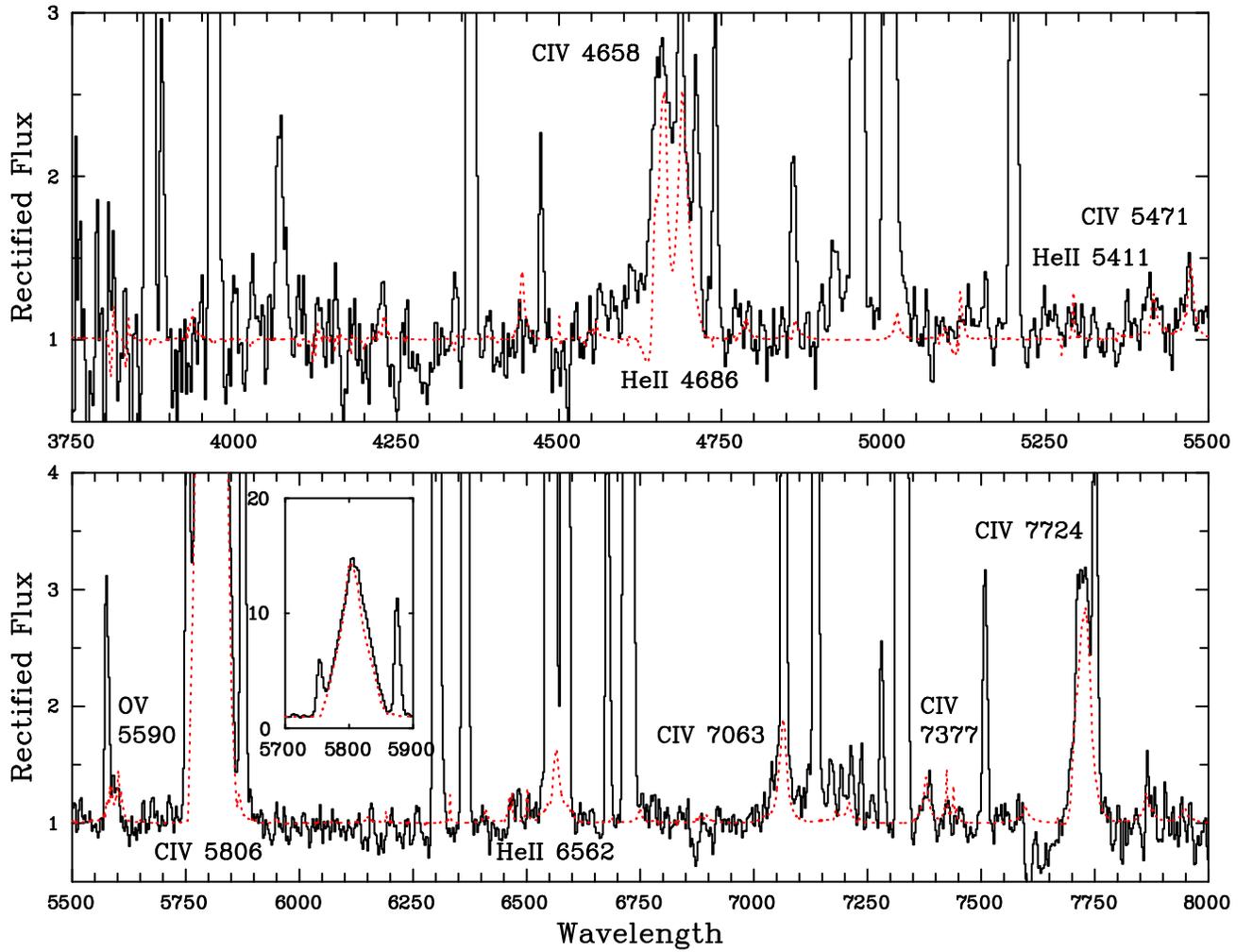}
\caption{A new ESO VLT/FORS spectrum obtained in 2001 June of V605 Aql (black line). The spectrum has been normalized. Overplotted in red is the best-fit non-LTE radiative transfer model atmosphere with the abundances in Table 1. Only the broad stellar emission lines have been fit. The narrow emission lines are from the surrounding PN, A58.}
\end{figure}

\end{document}